\documentclass[12pt]{article}
\usepackage{amsmath, amsthm, amsfonts, bm}
\usepackage{amssymb}
\usepackage{mathtools}
\usepackage{float}

\usepackage{hyperref}
\hypersetup{colorlinks=true, citecolor=blue, linkcolor=blue, filecolor=blue}
\usepackage{graphicx,psfrag,epsf, float}
\usepackage{enumerate,titlesec, color}
\usepackage{natbib}
\usepackage{footnote}
\usepackage{url} 
\usepackage{soul}
\usepackage{tikz}
\usepackage[ruled,linesnumbered]{algorithm2e}
\usepackage[font=small,labelfont=bf]{caption}

\usepackage{booktabs}
\usepackage{lineno}
\newcommand\revise[1]{{\color{black} #1}}
\usepackage{soul}

\usepackage{setspace}

\usepackage[textwidth=6.5in,bottom=1.8in,top=1in]{geometry}

\renewcommand\footnotemark{}

\usepackage{booktabs}

\usepackage{array}

\usepackage{multirow}
\definecolor{napiergreen}{rgb}{0.16, 0.5, 0.0}
\definecolor{myrtle}{rgb}{0.13, 0.26, 0.12}
\definecolor{dartmouthgreen}{rgb}{0.05, 0.5, 0.06}

\def\bfs{\mathbf s}
\def\bfv{\mathbf v}

\def\bfI{\mathbf I}
\def\bfX{\mathbf X}

\def\bfZ{\mathbf Z}
\def\bfD{\mathbf D}

\def\bfU{\mathbf U}

\def\bfalpha{\boldsymbol \alpha}
\def\bfbeta{\boldsymbol \beta}
\def\bfkappa{\boldsymbol \kappa}

\def\bfzeta{\boldsymbol \zeta}

\def\bfgamma{\boldsymbol \gamma}

\def\bfTheta{\boldsymbol \Theta}
\def\bfmu{\boldsymbol\mu}
\def\bflambda{\boldsymbol\lambda}

\def\bfmu{\boldsymbol\mu}

\def\bfzero{\boldsymbol 0}

\begin{document}
	\bigskip
	\date{}

\title{Bayesian semi-parametric inference for clustered recurrent events with zero-inflation and a terminal event}
\author{Xinyuan Tian, Maria Ciarleglio, Jiachen Cai, Erich J. Greene, \\ Denise Esserman,
Fan Li$^*$ and {Yize Zhao$^*$}\thanks{*Co-senior authors, and correspondence should be directed to: Fan Li (fan.f.li@yale.edu) and Yize Zhao (yize.zhao@yale.edu), Department of Biostatistics, Yale University, New Haven, CT 06511.}\bigskip\\
Department of Biostatistics, Yale University, New Haven, CT
}
	\maketitle

\def\spacingset#1{\renewcommand{\baselinestretch}%
	{#1}\small\normalsize} \spacingset{1}

 \vspace{-0.3cm}
\begin{abstract}
Recurrent event data are common in clinical studies when participants are followed longitudinally, and are often subject to a terminal event. With the increasing popularity of large pragmatic trials with a heterogeneous source population, participants are often nested in clinics and can be either susceptible or structurally unsusceptible to the recurrent process. These complications require new modeling strategies to accommodate potential zero-event inflation as well as hierarchical data structures in both the terminal and non-terminal event processes. In this paper, we develop a Bayesian semi-parametric model to jointly characterize the zero-inflated recurrent event process and the terminal event process. We use a point mass mixture of non-homogeneous Poisson processes to describe the recurrent intensity and introduce shared random effects from different sources to bridge the non-terminal and terminal event processes. To achieve robustness, we consider nonparametric Dirichlet processes to model the residual of the accelerated failure time model for the survival process as well as the cluster-specific frailty distribution, and develop a Markov Chain Monte Carlo algorithm for posterior inference. We demonstrate the superiority of our proposed model compared with competing models via simulations and apply our method to a pragmatic cluster randomized trial for fall injury prevention among the elderly.
\end{abstract}
\vspace{1cm}
\noindent%
{\textbf{Keywords}: Accelerated failure time model; Bayesian survival analysis; Dirichlet process; Pragmatic clinical trials; Semi-competing risks; Zero-inflation}

\spacingset{1.45}
\section{Introduction}
\label{s:intro}
Recurrent event data are common in clinical studies when participants are followed up longitudinally. Typically, each event occurrence can be subject to right censoring as well as a competing terminal event, such as death. In large pragmatic clinical trials, the event processes are often observed across a heterogeneous population, along with an informative competing event process subject to between-participant clustering. These features bring new challenges for the analysis of clustered recurrent events, due to the need for simultaneously characterizing the recurrent event process, non-terminal as well as terminal event survival process as a function of covariates.

\begin{figure}
\centering
\includegraphics[width=0.7\textwidth]{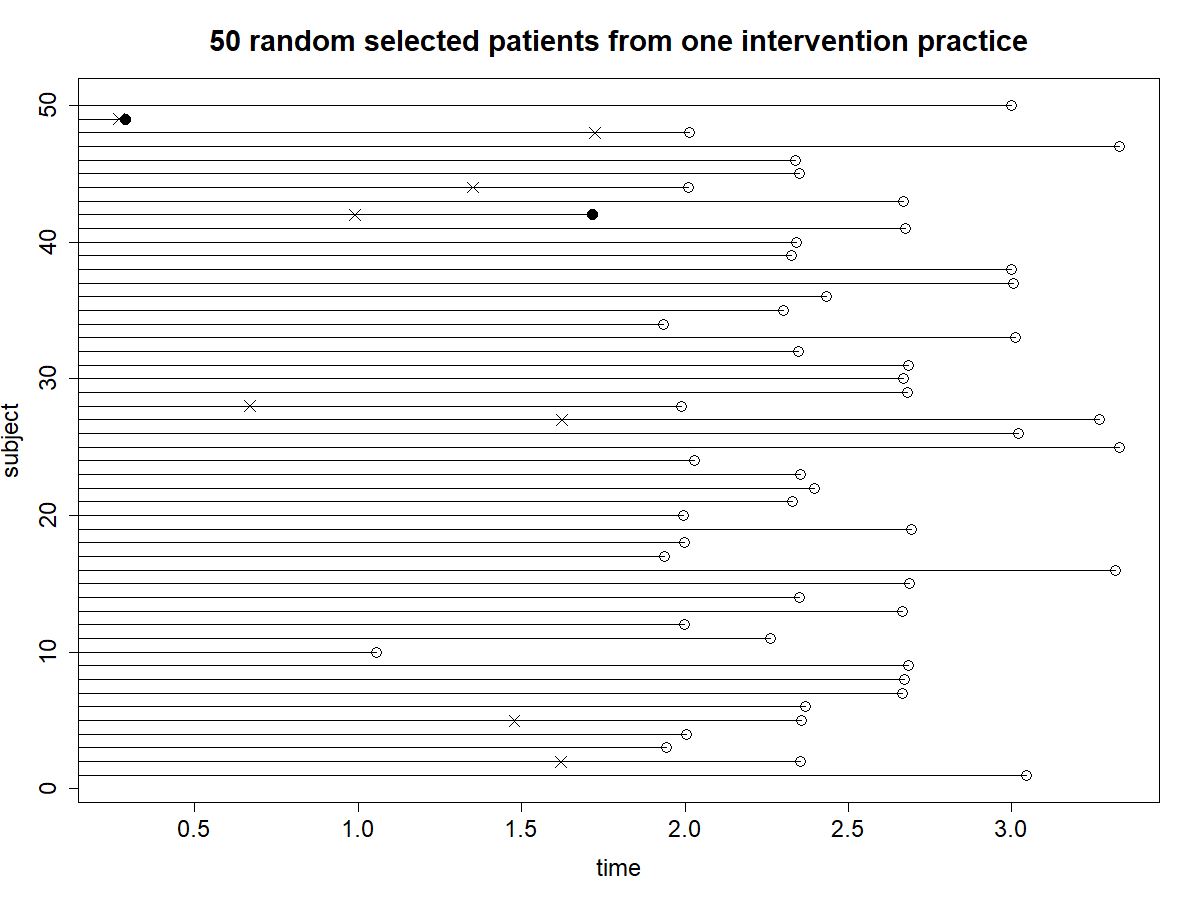}
\includegraphics[width=0.7\textwidth]{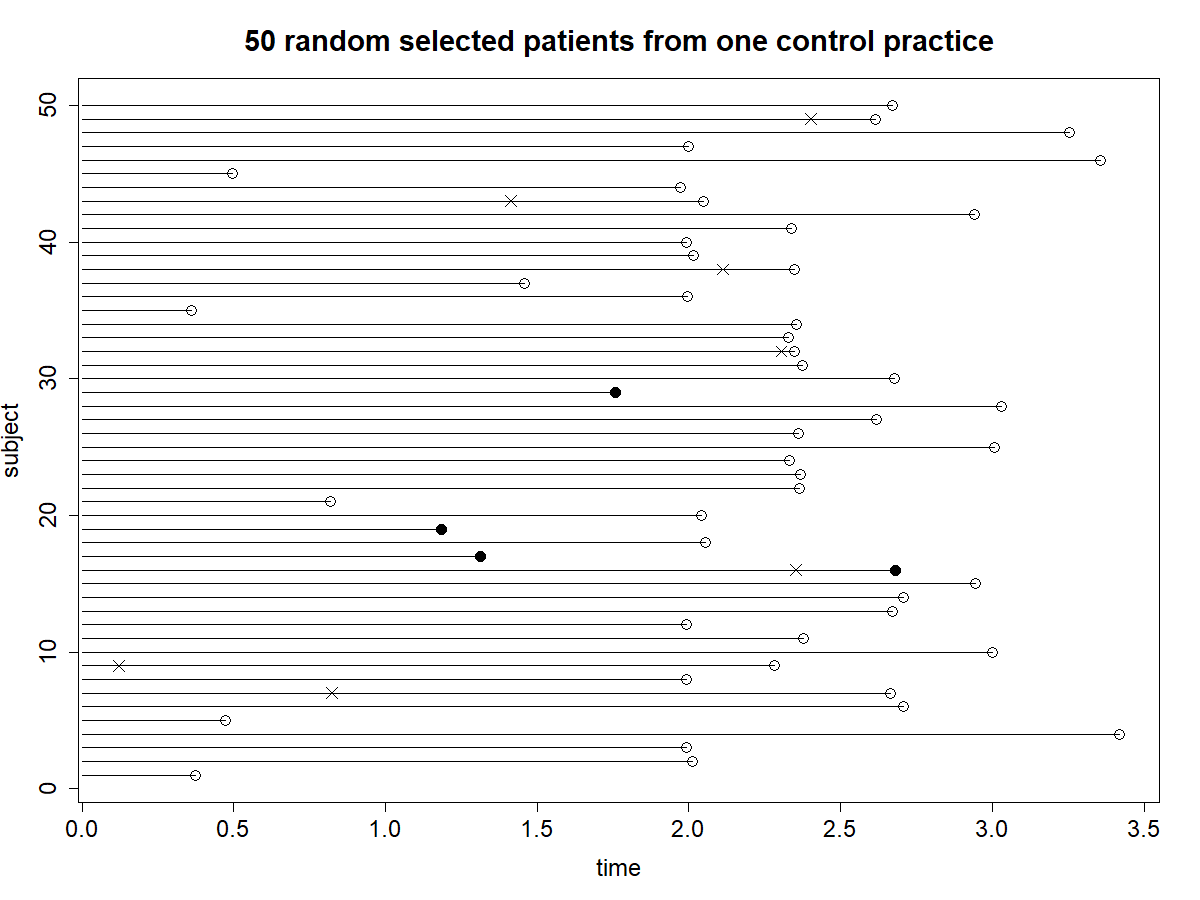}
     \caption{An illustration of (right-continuous) time trajectories for serious fall injury occurrence and terminal death event among randomly selected participants from both the intervention and control practices, where `$\circ$' represents censoring, `$\bullet$' represents occurrence of death, and `$\times$' represents the an occurrence of fall injury. }\label{stride}
\end{figure}

\textcolor{black}{Falls are the leading cause of injury-related death among older Americans, and approximate 1 in 4 older adults experiences fall each year, resulting in numerous deaths and injury related hospitalization and health care utilization annually \citep{verma2016falls,choi2019fall}. There has been a rising interest in implementing effective fall prevention strategies at a health care system level or provider level, to improve patient outcomes and reduce fall injury related mortality \citep{hopewell2018multifactorial}. In 2014, the Patient-Centered Outcomes Research Institute and the National Institute on Aging in the United States funded a pragmatic trial, the Strategies to Reduce Injuries and Develop Confidence in Elders \citep[STRIDE;][]{stride_introduction} study, to assess the effectiveness of a patient-centered intervention on fall injury prevention for older adults; our work is directly motivated by the STRIDE study. In STRIDE, more than 6,000 community-dwelling adults from 86 primary care practices were recruited, with 43 practices randomized to intervention and the remaining to usual care. Participants were followed up every four months via tele-interview (this is a relatively large number of clusters, as the upper quantile of number of clusters in a past systematic review by \citet{ivers2011impact} was only $52$)}. All reported fall injuries were recorded, and a blinded adjudication committee confirmed serious fall injures via medical and claim records from the participating healthcare systems and Centers for Medicare and Medicaid Services data \citep{Ganz2019}. During the study, 89\% of participants did not experience an adjudicated serious fall injury. \textcolor{black}{As a simple illustration, we randomly select $50$ patients from one random intervention practice and one random usual care practice, and present in Figure \ref{stride} the time trajectories for recurrent adjudicated serious fall injuries and an observed death event or censoring for each participant. Irrespective of the intervention, the recurrent event rate was relatively low, and there was an excessive number of participants without events, which signals potential zero-inflation for the recurrent event process. In addition, Web Figure 1 presents the descriptive Kaplan-Meier survival curves for the terminal event.}

There is a growing body of literature on the analysis of recurrent events in the presence of a terminal event. \textcolor{black}{For example, \citet{lancaster1998panel} represented the first effort to develop a recurrent event model with patient-level frailty subject to non-informative terminal events. \citet{sinha2008current} provided a comprehensive review of methods for recurrent event analysis with dependent termination and developed the first Bayesian approach to analyze such data. More recent developments for recurrent event analysis with dependent termination include} estimating equations approaches under a frequentist paradigm \citep{kalbfleisch2013estimating} and parametric or semi-parametric models under a Bayesian paradigm \citep{lin2017bayesian,li2019bayesian,li2020time}. A key feature of these methods is to characterize the dependence between non-terminal and terminal events under a semi-competing risk perspective \citep{fine2001semi}, as ignoring this dependence can lead to a biased inference. To do so, one common strategy is to formulate a joint model with a shared participant-level frailty in the recurrent event and terminal event submodels, where the submodels can either be based on the intensity functions of the event processes \citep{liu2004shared,lee_2019} or the hazard rate of the gap time between two events \citep{zhangsheng2011joint,paulon2020joint}. \textcolor{black}{Alternatively, \citet{xu2021joint} developed a joint latent class models to allow for class-specific risks for recurrence and termination. Their approach bypasses the distributional assumption of the shared random effect and can potentially lead to more interpretable covariate effects within and across latent classes.} Despite this growing literature, few existing methods have simultaneously addressed the complications of cluster correlated data featured in the STRIDE study, whereas failure to account for clustering can result in an invalid inference \citep{lee_2016}. 
\revise{\cite{jung2019joint} developed an approach that accounted for between-participant clustering in the presence of recurrent and terminal events. A similar joint model was also formulated in \cite{rondeau2015joint} and implemented in the \texttt{R} package \texttt{frailtypack}. However, these existing approaches require strong parametric assumptions on the between-participant clustering effect 
and have not accounted for population heterogeneity with respect to event susceptibility.}

The contributions of our work are several-fold. First, we propose a new joint model to analyze recurrent event and survival processes in the presence of between-participant clustering and a competing terminal event. We introduce random effects at the participant level and the practice level, both of which contribute to connecting the recurrent event and survival processes. Second, we address potential zero-inflation within our modeling framework by including a point mass at zero for the recurrent event intensity function. Using a latent indicator to define the status of unsusceptibility for each participant, we are able to directly inform population heterogeneity by separating the unobserved unsusceptible sub-population from the whole study population \citep{liu2016joint,kim2021joint}. \revise{Third, we consider separate nonparametric Dirichlet process priors \citep{ferguson1973bayesian} for the residual in the survival process as well as for the cluster-specific random effect, which, compared with conventional parametric formulations, alleviates potential bias due to model misspecification.} 
Finally, we apply the proposed Bayesian semi-parametric approach to analyze participant-level data from the STRIDE trial and generate new insights. 

The rest of the article is organized as follows. In Sections \ref{s:model} and \ref{sec:BA}, we introduce our Bayesian semi-parametric model including specifications of all submodels, choice of priors, and posterior inference. We evaluate the model performance by comparing with other competing approaches using simulations in Section \ref{sec:simu}. We provide a comprehensive analysis of  the STRIDE study in Section \ref{sec:data} using the proposed model and several other existing modeling techniques. We conclude with a discussion in Section \ref{s:discuss}.

\section{Modeling Clustered Recurrent Events in the Presence of a Terminal Event}\label{s:model}

We consider a clustered data structure
with recurrent events that are subject to a terminal event, such as death. We assume $J$ clusters (primary care practices) are recruited, with $N_j$ participants in cluster $j$ and $N=\sum_{j=1}^J N_j$ participants in total. 
Define $Q_{ij}(t)$ as the number of recurrent events prior to or at time $t$ for participant $i$ ($i=1,\ldots,N_j$) within cluster $j$ ($j=1,\ldots,J$). In our motivating STRIDE study, adjudicated serious fall injuries for patients are considered as recurrent events, subject to the risk of death as a terminal event. After a terminal event, recurrent events are no longer observable. In this case, we define time to the terminal event for each participant as $R_{ij}$ and the usual right censoring time (such as administrative censoring) as $C_{ij}$. The observed follow-up time is $\widetilde{R}_{ij}=R_{ij}\wedge C_{ij}$ with a censoring indicator $\Delta_{ij}=1$ if the terminal event is observed and 0 if censored. Equivalently, we observe a total of $Q_{ij}(\widetilde{R}_{ij})$ recurrent events for participant $i$ in cluster $j$. We also write $T_{ijk}\leq \widetilde{R}_{ij}$ as the time when the $k$th ($1\leq k\leq Q_{ij}(\widetilde{R}_{ij})$) recurrent event is observed. For notation purposes, we define the collection of recurrent event times for each participant with at least one event as $\mathbf{T}_{ij}=\left\{T_{ij1},\ldots,T_{ij,Q_{ij}(\widetilde{R}_{ij})}\right\}$, and for those with zero events as $\mathbf{T}_{ij}=\emptyset$. 




As shown in Figure \ref{stride}, a substantial proportion of participants in our motivating study have not experienced recurrent events, suggesting that some patients may be structurally unsusceptible to fall injuries during the study period, and could have distinctive characteristics  from the remaining population. This requires us to separately consider this subgroup for plausibly uncovering the actual event mechanisms. 
\revise{To model zero-inflation, we introduce a latent indicator $D_{ij}$ with $D_{ij}=1$ if participant $i$ in cluster $j$ belongs to the subgroup that is unsusceptible to recurrent event during the study period and 0 otherwise}. We consider a point mass mixture of non-homogeneous Poisson process (NHPP) to model the recurrent event hazard (or intensity) function for each participant as
\begin{equation}\label{eq:point}
  \lambda_{ij}(t) =
    \begin{cases}
      \gamma_{ij}\lambda_0(t)\exp{\left(\bfbeta^T\bfX_{ij}+\mu_j\right)} & \text{if $D_{ij}=0$;}\\
      0 & \text{if $D_{ij}=1$,}
    \end{cases}       
\end{equation}
In the hazard function \eqref{eq:point}, $\bfX_{ij}$ represent the covariates including the treatment arm and additional baseline characteristics potentially related to the recurrent process, $\bfbeta$ are the coefficients representing the relationship between $\bfX_{ij}$ and recurrent event process among the susceptible subgroup with $D_{ij}=0$, and $\lambda_0(t)$ is the associated baseline hazard. \revise{By definition, a participant belongs to the susceptible subgroup if $Q_{ij}(\widetilde{R}_{ij})>0$; otherwise the participant can belong to either the susceptible or unsusceptible subgroup.} In addition, $\gamma_{ij}$ is the subject-specific frailty accounting for the correlation between recurrent events for the same participant, and $\mu_j$ is the cluster-specific random effect that captures between-participant correlation within the same practice.

There are different options to specify $\lambda_0(t)$ under a Bayesian paradigm. For instance, we could assume a power law model \citep{lee_2019} with $\lambda_0(t)=\psi t^{\psi-1}$, which corresponds to a Weibull baseline hazard with scale parameter 1 and shape parameter $\psi$ controlling the rate of event occurrences. Alternatively, we could also consider a nonparametric specification for $\lambda_0(t)$ with a piecewise constant function \citep{mckeague2000bayesian,jung2019joint} 
\begin{equation}\label{eq:point2}
\lambda_0(t)= \sum_{g=1}^G \bfI\left[ s_{g-1}  < t  \leq s_{g}\right] \cdot\lambda_{0g},
\end{equation}
where $\bfI[\cdot]$ is the indicator function, $s_{0}=0$, $s_{G}$ represent the largest recurrent event time, and $\{s_{1},\ldots, s_{G-1}\}$ are $G-1$ grid points that partition the time interval such that baseline hazard is a constant $\lambda_{0g}$ over $(s_{g-1}, s_{g}]$. While the power law model assumes a monotone baseline hazard, the piecewise constant model can be more flexible and more robust to model assumptions. \textcolor{black}{In what follows, we will primarily focus on the piecewise constant model \eqref{eq:point2} as it tends to be more flexible; additional details and numerical results under the power law baseline hazard are provided in the Supplementary Materials (Web Appendix S2).}


For the survival process of the terminal event, we consider an accelerated failure time (AFT) model incorporating the hierarchical random effects shared with the recurrent event model 
\begin{equation}\label{eq:aft}
\log(R_{ij})=\alpha_0+\bfalpha^T\bfZ_{ij}+\xi_1\log(\gamma_{ij})+\xi_2\mu_j+\kappa_{ij}^{-1}\epsilon_{ij},
\end{equation}
where $\kappa_{ij}^{-1}$ is the participant-specific shape parameter and $\epsilon_{ij}$ is the independent and
identically distributed residual error for the log survival time. In \eqref{eq:aft}, $\alpha_0$ is the intercept that captures the common factor across subjects, $\bfZ_{ij}$ is the set of covariates associated with the terminal event time with coefficients $\bfalpha$ and can differ from $\bfX_{ij}$ in the recurrent event model \eqref{eq:point}, and coefficients $\xi_1$ and $\xi_2$ control the degree of unobserved associations between the recurrent and terminal event processes at the participant level and cluster level, respectively. This above model representation indicates that the participant-level frailty $\gamma_{ij}$ and the cluster-level random effect $\mu_j$ jointly affect the relative change in survival time for the terminal event to account for the variation beyond that captured by the observed covariates. Meanwhile, model \eqref{eq:aft} and the recurrent event intensity model \eqref{eq:point} share the hierarchical random effects to induce an informative terminal event process. To interpret this in the STRIDE study, an elderly participant who is more susceptible to repeated occurrences of falls may be either more likely or unlikely to survive until the end of the study, as captured by the participant-level frailty $\gamma_{ij}$ and its coefficient $\xi_1$. Similar interpretation also applies to the practice-level frailty $\mu_j$ and its coefficient $\xi_2$ in the terminal process submodel.

For AFT model \eqref{eq:aft}, a canonical parametric specification is to assume that residual error $\epsilon_{ij}$ follows a standard extreme value distribution and $\kappa_{ij}=\kappa$, $\forall~i,j$. Under this parameterization, the AFT model implies a Weibull hazard function for the terminal event time with $h_{ij}(t\mid \kappa_{ij}=\kappa)=\gamma_{ij}^{-\kappa\xi_1} t^{\kappa-1}\kappa\exp\left\{-\kappa\left(\alpha_0+\bfalpha^T\bfZ_{ij}+\xi_2\mu_j\right)\right\}$. Accordingly, the survival function becomes $H_{ij}(t\mid \kappa_{ij}=\kappa)=\exp\left[-\gamma_{ij}^{-\kappa\xi_1}t^{\kappa}\exp\big\{-\kappa\left(\alpha_0+\bfalpha^T\bfZ_{ij}+\xi_2\mu_j\right)\big\}\right]$. Such an AFT model with a homogeneous error distribution, although easy to implement, may be less robust to between-participant heterogeneity in their baseline risk to the terminal event. To enhance model robustness, we consider a nonparametric Dirichlet process ($\mathcal{DP}$) to model the error distribution. Specifically, we assume the participant-specific shape parameter
\begin{equation}\label{eq:DP0}
\kappa_{ij}\ \mid F \overset{\text{i.i.d}}{\sim} F,~~j=1,\dots, J; i=1,\dots, N_j; \qquad F\sim \mathcal{DP}(\phi_0, F_0).
\end{equation}
Here, $F_0$ is called a base measure that defines the expectation of the random probability $F\in \mathbb{R}$ from which $\kappa_{ij}$ is sampled, and $\phi_0$ is the scale parameter describing the overall sampling concentration or the variance of the random probability measure. We specify $F_0$ as a Gamma distribution $\mathcal{G}(a_{\kappa}, b_{\kappa})$, and assign a weakly-informative Gamma distribution for scale parameter $\phi_0\sim \mathcal{G}(1, 1)$ to ensure adequate flexibility. Essentially, model \eqref{eq:DP0} induces a nonparametric realization for the shape parameters, which then corresponds to a more flexible form of the hazard and survival functions. To elaborate on this point, we can represent the $\mathcal{DP}$ model in \eqref{eq:DP0} by an infinite mixture of point masses \citep{sethuraman1994constructive} 
\begin{eqnarray}
F&=&\sum_{k=1}^\infty \pi_k\mu_{\theta_k}, \qquad \text{with} \quad \pi_k=\pi'_k\prod_{h=1}^{k-1}(1-\pi'_h),\label{DP3}
\end{eqnarray}
where $\mu_\theta$ is a probability measure concentrated at $\theta$, and the
two sets of independent and identically distributed random variables $\{\pi'_k\}_{k=1}^{\infty}$ and $\{\theta_k\}_{k=1}^{\infty}$ follow
\begin{eqnarray}\label{DP2}
\pi'_k\mid G_0,\phi \sim \text{Beta}(1, \phi_0); \qquad \theta_k\mid F_0,\phi\sim F_0;\qquad k=1,\dots,\infty.
\end{eqnarray}
Here, $\{\theta_k\}_{k=1}^\infty$ is a sequence of independent draws from the base measure $F_0$, and $\{\pi_k\}_{k=1}^\infty$ are the weight parameters constructed via a stick-breaking representation. With probability one, $F$ is a discrete distribution as a combination of infinite number of point masses. Under weights $\{\pi_k\}_{k=1}^\infty$, realization of each $\kappa_{ij}$ will be obtained directly from $F$ consisting of components $\{\theta_k\}_{k=1}^\infty$. The induced survival function for participant $i$ in cluster $j$ then becomes an infinite mixture of Weibull survival functions given by
\begin{equation*}
\sum_{k=1}^\infty \pi_k H_{ij}(t\mid \kappa_{ij}=\theta_k)=\sum_{k=1}^\infty \pi_k \exp\left[-\gamma_{ij}^{-\theta_k\xi_1}t^{\theta_k}\exp\big\{-\theta_k\left(\alpha_0+\bfalpha^T\bfZ_{ij}+\xi_2\mu_j\right)\big\}\right],
\end{equation*}
and the associated hazard function corresponds to a similar infinite mixture of Weibull hazards
\begin{align*}
\sum_{k=1}^\infty \left\{\frac{\pi_k H_{ij}(t\mid \kappa_{ij}=\theta_k)}{\sum_{l=1}^\infty \pi_l H_{ij}(t\mid \kappa_{ij}=\theta_l)}\right\}
\gamma_{ij}^{-\theta_k\xi_1} t^{\theta_k-1}\theta_k\exp\left\{-\theta_k\left(\alpha_0+\bfalpha^T\bfZ_{ij}+\xi_2\mu_j\right)\right\},
\end{align*}
both of which are arguably much more flexible than their canonical, fully parametric counterparts. Meanwhile, as shown in \eqref{DP3}, with $k$ increased, $\pi_k$ decreases exponentially and concentrates the sampling on a number of initial components. This allows the residual error distributions to group based on their identical shape parameter values, and in turn, induces a clustering effect to dissect subgroup of individuals sharing a similar shape of the survival function. Finally, the canonical AFT specification can be considered as a special case of \eqref{DP3} with a degenerate Dirac measure. 

\section{Bayesian Inference}\label{sec:BA}
\subsection{Prior Specification}\label{sec:prior}
To jointly characterize the zero-inflated recurrent events and terminal event process, the proposed joint modeling framework involves the following unknown parameters:
regression coefficients $\bfbeta$, $\bfalpha$, $\xi_1$, and $\xi_2$, participant-level and cluster-level random effects $\bfgamma=\{\gamma_{ij}\}$ and $\bfmu=\{\mu_j\}$, latent indicator $\bfD=\{D_{ij}\}$, \revise{shape parameter $\bfkappa=\{\kappa_{ij}\}$ for the terminal event submodel, grid points $\bfs$, and piecewise constants $\bflambda=(\lambda_{01},\dots,\lambda_{0G})$ for the recurrent event submodel. } 

The hierarchical random effects play an important role in connecting the recurrent and survival processes, since they represent shared unmeasured factors in addition to those captured by the baseline covariates. 
The frailty $\bfgamma$ is directly grouped by different practices the participants belong to and provides quantification of between-participant heterogeneity, while the practice-specific random effects $\bfmu$ account for between-practice heterogeneity. We assume independence between elements of $\bfgamma$ and assign $\gamma_{ij}\sim \mathcal{LN}(0,\tau_j^2)$, where $\mathcal{LN}$ represents a log-normal distribution and $\tau^2_j$ represents a practice-specific variance parameter; and we adopt an Inverse Gamma ($\mathcal{IG}$) hyper-prior such that $\tau_j^2\sim \mathcal{IG}(a_0,b_0)$. For $\bfmu$, instead of using parametric conjugate priors, we consider a nonparametric $\mathcal{DP}$ prior by assuming
\begin{equation}\label{eq:DP}
\mu_j\ \mid G \overset{\text{i.i.d}}{\sim} G,~~j=1,\dots, J; \qquad G\sim \mathcal{DP}(\phi, G_0).
\end{equation}
We specify base measure $G_0$ as a Normal distribution, $\mathcal{N}(0, \sigma^2)$, and assign $\phi\sim \mathcal{G}(1, 1)$ to ensure adequate flexibility. Prior \eqref{eq:DP} induces a nonparametric representation for the random effects over practices. Since the inference of model parameters may be sensitive to parametric assumptions of the practice-level random effects \citep{gasparini2019impact}, this nonparametric prior can induce more robust characterization of the quality of care in each practice. To facilitate posterior inference under \eqref{eq:DP}, following \eqref{DP3} and \eqref{DP2}, we also resort to an infinite mixtures of point masses representation under the point mass random set $\{\eta_{l}\}_{l=1}^{\infty}$ where each $\mu_j$ sampled from under the weights $\{\tilde{\pi}_{l}\}_{l=1}^{\infty}$ ($\eta$ and $\tilde{\pi}$ are analogs to those introduced in model \eqref{DP3}). This also groups  realizations of each element within $\bfmu$ together by their identical values, indicating, for example, similar quality of care across the included practices. In STRIDE, the primary care practices are nested within different health care systems, which could induce  inter-practice similarity. Although we do not directly account for heterogeneity across health systems beyond that across practices, the implicit clustering effect due to the $\mathcal{DP}$ prior automatically identifies more similar practices according to values of $\mu_j$, either within or across different health systems, and provides additional flexibility beyond a single random effect at the health system level. Alternatively, in the absence of a clear grouping pattern between practices in terms of quality of care, we can still rely on \eqref{eq:DP} to potentially reduce the number of unknown practice-level random effects. \textcolor{black}{Of note, in the analysis of the STRIDE trial, we have specified the above log-normal parametric prior for the patient-level frailty because the recurrent event rate was relatively low; however, a relatively large number of practices in STRIDE supports a nonparametric $\mathcal{DP}$ prior for the practice-level random effects. In addition, we have considered a $\mathcal{DP}$ shape-mixture of errors in the AFT terminal event model as well as a $\mathcal{DP}$ prior for the practice-level random effects in the same terminal event model. This double nonparametric prior specification does not lead to non-identifiability because the practice-level random effects are shared between the recurrent event and terminal event models and posterior inference for the practice-level random effects will be based on additional information beyond the terminal event process.}

\begin{figure}
    \centering
   \includegraphics[width=0.9\textwidth]{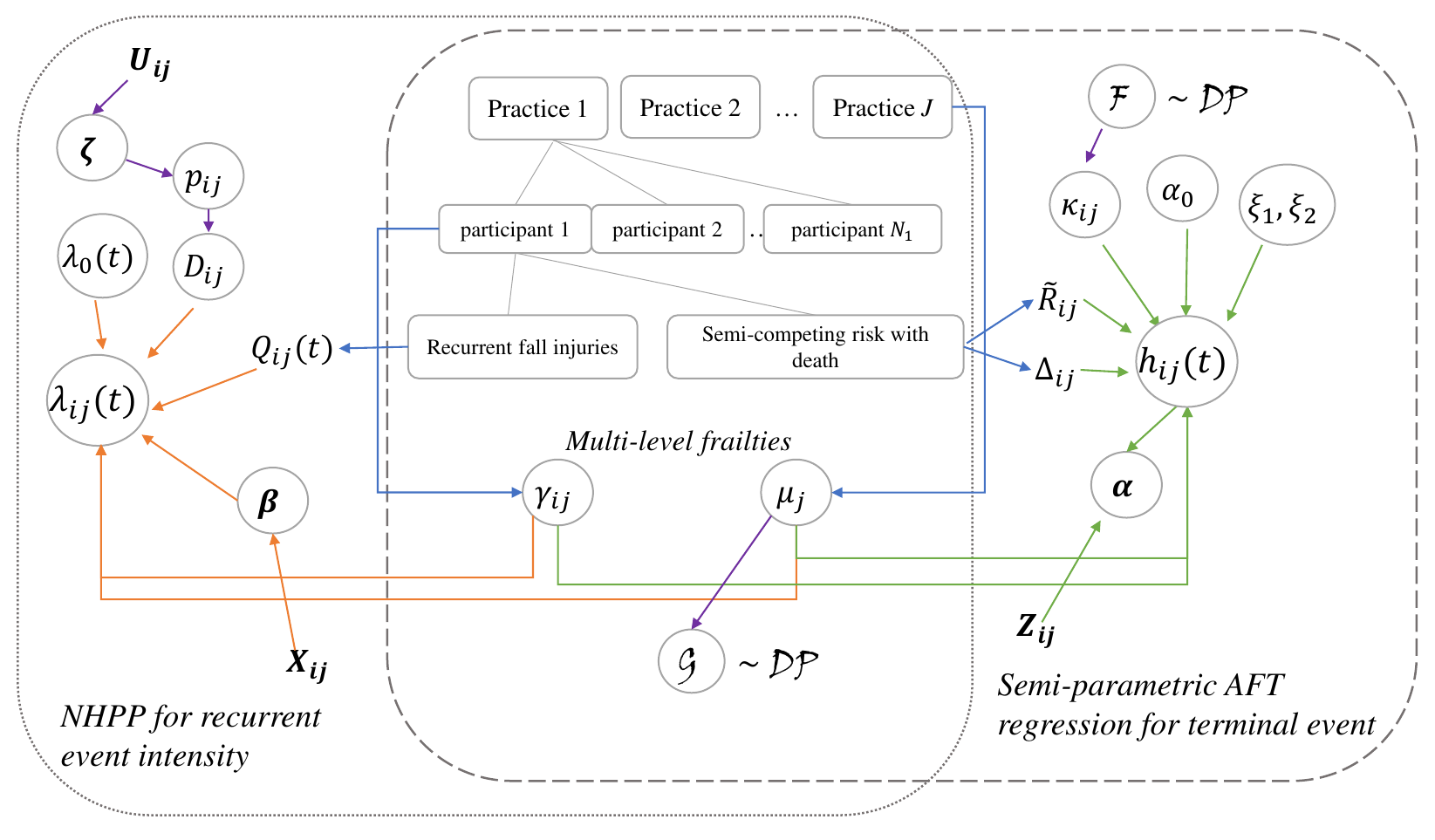}
    \caption{A graphical demonstration of the multi-level data structure and our proposed Bayesian joint model for the zero-inflated recurrent events and semi-competing survival process. }
    \label{fig:demo}
\end{figure}

We further assume the latent indicator $D_{ij}$ follows $D_{ij}\sim \text{Bern}(p_{ij})$, $i=1,\dots,N_j$,  $j=1,\dots,J$, with $p_{ij}$ being the participant-specific probability to be classified into the unsusceptible subgroup. In practice, when there is prior knowledge on potential risk factors that are associated with an individual's susceptibility status for recurrent events, we can adopt a logistic model 
\begin{eqnarray}\label{eq:logit}
\text{logit}(p_{ij})=\bfzeta^T\bfU_{ij},
\end{eqnarray}
where $\bfU_{ij}$ includes an intercept as well as risk factors for susceptibility and $\bfzeta$ represents the regression coefficients \citep{Berkson&Gage,Cooner}. In other cases without strong prior information on such covariates, we could instead assume that $p_{ij}$ takes a constant value, say $0.5$, which leads to a non-informative prior for the latent indicator $D_{ij}$, and can be regarded as a special case of \eqref{eq:logit}.
For generality, we will discuss posterior inference under a general logistic formulation \eqref{eq:logit}. \textcolor{black}{In terms of the baseline hazard in the recurrent event process, following \cite{jung2019joint}, we pre-specify $\bfs$ as quantiles based on the minimum to the maximum recurrent event time, and adopt a uniform prior $(0, \infty)$ for each element within $\bflambda$, i.e, $p(\lambda_{0g})\propto 1$. This improper uniform prior is a convenient choice and still leads to well-defined posterior distribution that integrates to 1, and results under a proper uniform prior over $(0,100)$ are no different for the analysis of STRIDE (omitted for brevity).} To complete prior specification, we assign priors for the remaining model parameters such that $\bfbeta\sim \mathcal{N}(\bfzero,\sigma^2_\beta\mathbf{I})$, $\bfalpha\sim \mathcal{N}(\bfzero,\sigma^2_\alpha\mathbf{I})$, $\bfzeta\sim \mathcal{N}(\bfzero,\sigma^2_\zeta\mathbf{I})$, $\xi_1\sim \mathcal{N}(0,\sigma^2_{\xi_1})$, $\xi_2\sim \mathcal{N}(0,\sigma^2_{\xi_2})$; and further assign conjugate $\mathcal{IG}$ hyper-priors for $\sigma_{\beta}^2$ and $\sigma_{\alpha}^2$ and \textcolor{black}{pre-specify the remaining hyper-parameters with reasonable values without strong prior impact on the posterior inference; as our model includes a substantial amount of parameters, sensitivity analyses to choice of hyper-parameters are also recommended.} For an overview of our method, Figure \ref{fig:demo} provides a graphical illustration of the data structure along with key modeling assumptions.

\subsection{Likelihood and Posterior Inference}\label{sec:posterior}
Given the observed data $\mathcal{O}_{ij}=\left\{\widetilde{R}_{ij}, \Delta_{ij}, \mathbf{T}_{ij}, Q_{ij}(\widetilde{R}_{ij}), \bfX_{ij}, \bfZ_{ij}, \bfU_{ij}\right\}$ for each subject $i~ (i=1,\dots, N_j)$  within practice $j~ (j=1,\dots,J)$, we need to carefully distinguish between different events and survival states, as well as the subgroup each subject belongs to, in order to carry out inference for all model parameters. \revise{For example, while each participant may be \textcolor{black}{or} may not be susceptible to recurrent events, all participants are susceptible to the terminal events in the STRIDE application.} 
With the unknown parameters $\bfTheta=\left\{\bfbeta,\bfalpha,\bfzeta,\bfgamma,\bfmu,\bfD,\xi_1,\xi_2,\bfkappa; \phi \text{ or } \bflambda \right\}$,  the observed data likelihood involves a combination of probabilities for structural zeros among the unsusceptible subgroup, recurrent events, and terminal events, and is given by
\revise{\begin{eqnarray*}\label{eq:likelihood}
\mathcal{L}(\{\mathcal{O}_{ij}\}\mid \bfTheta)&=& \prod^J_{j=1}\prod_{i=1}^{N_j}\left\{D_{ij}+(1-D_{ij})(1-\Delta_{ij})S_{ij}(\widetilde{R}_{ij})H_{ij}(\widetilde{R}_{ij})+(1-D_{ij})\Delta_{ij}S_{ij}(\widetilde{R}_{ij})\right.\\
&&\left.\times f_{ij}(\widetilde{R}_{ij})\right\}^{\bfI\left[Q_{ij}(\widetilde{R}_{ij})=0\right]}\times
\left\{\Delta_{ij}\prod_{k=1}^{Q_{ij}(\widetilde{R}_{ij})}\lambda_{ij}(T_{ijk})S_{ij}(\widetilde{R}_{ij})f_{ij}(\widetilde{R}_{ij})\right.\\
&&\left.+(1-\Delta_{ij})\prod_{k=1}^{Q_{ij}(\widetilde{R}_{ij})}\lambda_{ij}(T_{ijk})S_{ij}(\widetilde{R}_{ij})H_{ij}(\widetilde{R}_{ij})\right\}^{\bfI\left[Q_{ij}(\widetilde{R}_{ij})>0\right]},
\end{eqnarray*}}where $f_{ij}(\widetilde{R}_{ij})=h_{ij}(\widetilde{R}_{ij})H_{ij}(\widetilde{R}_{ij})$ is the density function for the terminal event process of participant $i$ in practice $j$ evaluated at the observed survival time $\widetilde{R}_{ij}$, and the indicator function $\bfI[\cdot]$ separating the likelihood for those with and without recurrent events. By combining the observed data likelihood with our prior specification, we obtain the joint posterior distribution of $\bfTheta$, from which we perform estimation and inference for each of the unknown parameters.

To achieve posterior inference, we develop a Markov Chain Monte Carlo (MCMC) algorithm based on a combination of Gibbs sampler and Metropolis-Hastings (MH) updates. The full computational details of our MCMC are provided in the Supplementary Materials. In brief, under random initials, the algorithm cycles through the following steps:
\begin{itemize}
  \setlength\itemsep{0.2em}
\item Sample each element of $\bfD$ from its posterior Bernoulli distribution.
\item For the recurrent event submodel, update each element of $\bfbeta$ via its MH step; \revise{and update each element of $\bflambda$ in the baseline hazard from its MH step.} 
\item For the terminal event submodel, update $\alpha_0$ and each element of $\bfalpha$ via the corresponding MH steps. \revise{For the individual shape parameter $\bfkappa$,  we implement an approximate sampling procedure under the truncated stick-breaking process \citep{ishwaran2001gibbs,li2015spatial}, where a conservative upper bound $K$ larger than the possible number of latent groups for the mixture of $\kappa_{ij}$'s is assigned. 
By introducing a mapping indicator set $\bfv=(v_{11},\dots,v_{N_J,J})$ with $v_{ij}\in\{1,\dots,K\}$ following a Multinomial distribution with probabilities $\{\pi_1,\dots,\pi_K\}$, we align each $\kappa_{ij}$ to its latent membership label $v_{ij}$.  Within the same group membership label, the $\kappa_{ij}$'s are considered identical. Therefore, we update each $v_{ij}$ from the posterior Multinomial distribution, sample $\kappa_{ij}$ within each of the $K$ clusters via a MH step, and update $\pi_k=\pi'_k\prod_{h<k}(1-\pi'_h)$ with $\pi'_h$ sampled from the Beta distribution.}
\item  For the participant-specific frailty, update each $\gamma_{ij}$ via the MH step and sample the frailty variance $\tau_j^2$ from its posterior $\mathcal{IG}$ distribution for $i=1,\dots, N_j, j=1,\ldots,J$.
\item For the practice-specific random effect $\bfmu$, we implement a similar sampling procedure as that for $\bfkappa$ by assigning a conservative upper bound $L$ and
introducing a mapping indicator set $\mathbf{m}=(m_1,\dots,m_J)$ with each element following a Multinomial distribution with probabilities $\{\tilde{\pi}_1,\dots,\tilde{\pi}_L\}$. The update for each $\mu_j$, $m_j$ and $\tilde{\pi}_l$ follow a similar procedure to that used in updating the shape parameter of the terminal event submodel.
\item For the shared random effects in the terminal event submodel, update $\xi_1$ and $\xi_2$ via the corresponding MH steps.
\item For the logistic model, update each element of  $\bfzeta$ via the corresponding MH step.
\end{itemize}
 In our practice, we confirm the posterior convergence by both trace plots and the Gelman-Rubin method \citep{gelman1992inference}. Based on the posterior samples (after burn-in), we can directly obtain the point and credible interval estimators for each parameter using the posterior mean and associated quantiles.

\section{Simulation Studies} \label{sec:simu}

We carry out simulation studies to assess the finite-sample performance of the proposed Bayesian semi-parametric joint model and compare with alternative modeling approaches. Although our motivating STRIDE study recruited 86 practices, we simulate 60, 40, 20 practices, representing more challenging scenarios with fewer clusters. We assume equal numbers of participants per practice and consider $N=1,800$, $N=1,200$ and $N=600$ as three levels of total sample sizes. For each participant, we specify the covariates $\bfZ_{ij}$ for the terminal event  as a three-dimensional vector with each element generated from $\mathcal{N}(0,0.1^2)$ and set $\bfalpha=(0.2, 0.3, 0.4)^T$. We then generate the frailty $\gamma_{ij}\sim \mathcal{LN}(0, 0.3)$ with a common variance component across all practices, and we consider a five-component mixture of Normals to simulate the practice-level random effect, 
$$\mu_{j} \stackrel{\text{ind}}{\sim}0.2\mathcal{N}(-0.4,0.1^2)+0.2\mathcal{N}(-0.2,0.1^2)+0.2\mathcal{N}(0,0.1^2)+0.2\mathcal{N}(0.2,0.1^2)+0.2\mathcal{N}(0.4,0.1^2).$$ We simulate the survival time $R_{ij}$ from a mixture of Weibull distributions with the shape parameter $\kappa_{ij}$ drawing randomly from the discrete value set $\{0.7, 2.2, 5.2, 8.2\}$ with equal probability and $\alpha_0=0.15$, $\xi_1=0.1$, $\xi_2=-0.5$. To determine censoring status, we generate independently from $\Delta_{ij}{\sim}\text{Bern}(0.5)$. When $\Delta_{ij}=0$, we generate the observed survival time $\widetilde{R}_{ij}$ from a Uniform distribution under $(0, R_{ij})$; otherwise, we directly equate $\widetilde{R}_{ij}=R_{ij}$. For the recurrent process, we first specify the covariates $\bfX_{ij}$ for the recurrent events including the first two elements of $\bfZ_{ij}$ and a third element generated from $\mathcal{N}(0,0.1^2)$, then we set $\bfbeta=(0.4, 0.3, 0.2)^T$. We generate the latent indicator $D_{ij}$ from Bernoulli distribution with participant-specific probability  $p_{ij}=0.5$ to be classified into the unsusceptible subgroup. To generate the recurrent event process $Q_{ij}(t)$, we consider a piecewise constant baseline hazard specified by quintile grids and $(\lambda_{01},\lambda_{02},\lambda_{03},\lambda_{04},\lambda_{05})^T=(2,2.3,2.1,2.4,1.7)^T$ and censor the recurrent events at time $\widetilde{R}_{ij}$. For the unsusceptible subgroup, we set $Q_{ij}(\widetilde{R}_{ij})=0$ but without affecting the terminal event time.  \textcolor{black}{Besides the above data generating process (referred to as DGP1), to assess the robustness of our method, we also consider two additional scenarios where the data generations do not follow our model assumptions. Specifically, we first repeat the above data-generated setting but simulate the practice-level random effects from a single-component Normal distribution $\mathcal{N}(0,0.1^2)$ (referred to as DGP2), which represents a simpler case. Additionally, we also consider a scenario where individual-level heterogeneity exists for the terminal event process with shape parameter $\kappa_{ij}$ in the AFT model randomly generated from a Gamma distribution $\mathcal{G}(1,1)$ (referred to as DGP3).} Throughout, we consider three different sample sizes as introduced earlier and simulate $250$ data replicates for each setting.

\revise{To implement our method, we set $\sigma^2_{\zeta}=\sigma^2_{\xi_1}=\sigma^2_{\xi_2}=10$ to give non-informative Normal priors, $a_{\kappa}=b_{\kappa}=1$ for the Gamma base measure, $\sigma^2=1$ for the normal base measure, and $\mathcal{IG}(1/2, 1/2)$ for the conjugate priors of $\sigma^2_{\beta}$ and  $\sigma^2_{\alpha}$. We also consider $G=5$ to specify the quantile grids in $\bfs$. In each implementation, multiple chains with randomly generated initial values are run for 10,000 iterations
with the first 5,000 as burn-in. Our results show that the posterior inference is insensitive to the
initial values with a proper mixing for each parameter. In addition to implementing our proposed model (abbreviated as BMZ-$\mathcal{DP}$ for the Bayesian multi-level zero-inflated $\mathcal{DP}$ model), we also consider three variations of BMZ-$\mathcal{DP}$ by simplifying certain model components: (1) BM-$\mathcal{DP}$, which ignores the structural zeros by modeling recurrent event hazard with a single mode Poisson process; (2) BZ-$\mathcal{DP}$, which ignores the multi-level data structure by omitting the practice-level random effects; (3) BMZ, which replaces the nonparametric $\mathcal{DP}$ prior for $\mu_j$ with a fully parametric normal prior and the nonparametric $\mathcal{DP}$ prior for $\kappa_{ij}$ with a fully parametric gamma prior, as well as (4) the joint frailty model under a frequentist paradigm implemented in the {\tt{R}} package {\tt{frailtypack}} \citep{rondeau2012frailtypack,rondeau2015joint}, which accounts for the multi-level data structure but ignores structural zeros.} Of note, BZ-$\mathcal{DP}$ is a variation of the approach developed in \cite{lee_2019} with the addition of the susceptible subgroup, and BMZ is a pure Bayesian parametric implementation. The priors and hyper-parameters for the three Bayesian model variations largely follow those for BMZ-$\mathcal{DP}$; and for the frequentist joint frailty model (denoted as frailty), we use the {\tt{frailtyPenal}} function which is designed to fit a joint frailty model for clustered data and closest to our setting (with gamma-distributed participant-level frailty and practice-level frailty). For each method, we summarize the mean or posterior mean, percentage bias (\%) relative to the true value, and the 95\% confidence or credible intervals (CIs) for the primary parameters of interest, $\bfbeta=(\beta_1,\beta_2,\beta_3)^T$ and $\bfalpha=(\alpha_1,\alpha_2,\alpha_3)^T$ in Table \ref{table:result} under the original complex data generating process (DGP1). \textcolor{black}{The results under DGP2 (single-component normal practice-level random effects) and DGP3 (individual-specific shape parameter for the terminal event process) are summarized in Web Tables 2 and 3, respectively.}

\begin{table}
\centering
 \caption{Simulation results under sample sizes 600, 1200, 1800 for all the methods summarized by mean or posterior mean (Mean), percentage bias (Bias (\%)) and coverage probability of the 95\% credible interval (Coverage (\%)).}\label{table:result}
\resizebox{0.75\textwidth}{!}{
\begin{tabular}{cccrrrrrrr}
\toprule
                &                  &      & \multicolumn{3}{c}{Recurrent Process} &  & \multicolumn{3}{c}{Survival Process} \\
                \cmidrule(lr){4-6}\cmidrule(lr){7-10}
\multirow{3}{*}{} $N$ &                  &  Method    & $\beta_{1}$          & $\beta_{2}$         & $\beta_3$         &  & $\alpha_1$         & $\alpha_2$         & $\alpha_3$         \\
\midrule
                  &                  & BMZ-$\mathcal{DP}$ & 0.41        & 0.31       & 0.25       &  & 0.20       & 0.32       & 0.40       \\
                  &                  & BM-$\mathcal{DP}$ & -0.04       & -0.15      & -0.22      &  & 0.21       & 0.31       & 0.41       \\
\multirow{3}{*}{} & Mean & BZ-$\mathcal{DP}$ & 0.52        & 0.45       & 0.37       &  & 0.21       & 0.32       & 0.40       \\
                  &                  & BMZ & 0.43        & 0.35       & 0.29       &  & 0.22       & 0.33       & 0.42       \\
                  &                  & Frailty & -0.08       & -0.16      & -0.28      &  & 0.02       & 0.06       & -0.05      \\
                  \cmidrule(lr){4-6}\cmidrule(lr){7-10}
                  &                  & BMZ-$\mathcal{DP}$ & 2.20        & 2.93       & 26.22      &  & 2.21       & 6.25       & 0.67       \\
                  &                  & BM-$\mathcal{DP}$ & -111.15     & -148.38    & -210.02    &  & 4.63       & 2.85       & 2.57       \\
600               & Bias (\%)  & BZ-$\mathcal{DP}$ & 29.03       & 48.87      & 84.75      &  & 6.92       & 8.15       & 0.58       \\
                  &                  & BMZ & 9.40        & 17.41      & 44.09      &  & 10.72      & 9.69       & 3.83       \\
                  &                  & Frailty & -118.88     & -153.00    & -237.96    &  & -90.21     & -80.77     & -113.39    \\
                  \cmidrule(lr){4-6}\cmidrule(lr){7-10}
                  &                  & BMZ-$\mathcal{DP}$ & 88.59       & 84.56      & 79.19      &  & 95.91      & 92.87      & 96.95      \\
                  &                  & BM-$\mathcal{DP}$ & 1.49        & 1.33       & 1.33       &  & 93.81      & 90.90      & 75.36      \\
                  & Coverage  (\%)        & BZ-$\mathcal{DP}$ & 66.32       & 54.63      & 50.98      &  & 93.87      & 89.79      & 95.91      \\
                  &                  & BMZ & 82.31       & 79.02      & 70.95      &  & 91.83      & 85.37      & 89.79      \\
                  &                  & Frailty & 1.49        & 1.49       & 0.00       &  & 79.66      & 74.58      & 62.71      \\
                  \hline
                  &                  & BMZ-$\mathcal{DP}$ & 0.38        & 0.31       & 0.23       &  & 0.20       & 0.31       & 0.40       \\
                  &                  & BM-$\mathcal{DP}$ & -0.05       & -0.13      & -0.24      &  & 0.21       & 0.31       & 0.43       \\
                  & Mean & BZ-$\mathcal{DP}$ & 0.50        & 0.44       & 0.37       &  & 0.21       & 0.32       & 0.41       \\
                  &                  & BMZ & 0.42        & 0.35       & 0.27       &  & 0.21       & 0.33       & 0.41       \\
                  &                  & Frailty & -0.08       & -0.17      & -0.28      &  & 0.05       & -0.03      & -0.05      \\
                  \cmidrule(lr){4-6}\cmidrule(lr){7-10}
                  &                  & BMZ-$\mathcal{DP}$ & -4.66       & 2.93       & 16.37      &  & 0.27       & 3.52       & 0.96       \\
                  &                  & BM-$\mathcal{DP}$ & -111.81     & -144.49    & -220.16    &  & 4.59       & 3.43       & 6.38       \\
1200               & Bias (\%)  & BZ-$\mathcal{DP}$ & 25.30       & 46.01      & 84.45      &  & 4.03       & 6.89       & 1.77       \\
                  &                  & BMZ & -121.00     & 16.70      & 36.13      &  & 2.64       & 8.34       & 3.57       \\
                  &                  & Frailty & 6.18        & -158.67    & -238.52    &  & -77.35     & -108.81    & -113.37    \\
                  \cmidrule(lr){4-6}\cmidrule(lr){7-10}
                  &                  & BMZ-$\mathcal{DP}$ & 84.85       & 84.52      & 83.46      &  & 94.87      & 90.34      & 94.92      \\
                  &                  & BM-$\mathcal{DP}$ & 1.12        & 1.12       & 0.03       &  & 84.39      & 90.88      & 52.32      \\
                  & Coverage  (\%)        & BZ-$\mathcal{DP}$ & 58.86       & 40.56      & 25.34      &  & 86.77      & 85.13      & 95.46      \\
                  &                  & BMZ & 80.92       & 72.34      & 71.76      &  & 82.54      & 85.37      & 85.47      \\
                  &                  & Frailty & 1.33        & 1.33       & 1.33       &  & 71.19      & 62.71      & 55.93      \\
                  \hline
                  &                  & BMZ-$\mathcal{DP}$ & 0.37        & 0.31       & 0.22       &  & 0.20       & 0.30       & 0.40       \\
                  &                  & BM-$\mathcal{DP}$ & -0.05       & -0.13      & -0.24      &  & 0.21       & 0.31       & 0.42       \\
                  & Mean & BZ-$\mathcal{DP}$ & 0.51        & 0.44       & 0.35       &  & 0.22       & 0.32       & 0.40       \\
                  &                  & BMZ & 0.43        & 0.37       & 0.27       &  & 0.21       & 0.32       & 0.42       \\
                  &                  & Frailty & -0.07       & -0.16      & -0.27      &  & 0.05       & -0.01      & -0.04      \\
                  \cmidrule(lr){4-6}\cmidrule(lr){7-10}
                  &                  & BMZ-$\mathcal{DP}$ & -7.87       & 4.98       & 13.44      &  & 1.35       & 1.60       & 1.38       \\
                  &                  & BM-$\mathcal{DP}$ & -114.50     & -142.90    & -220.50    &  & 4.04       & 2.47       & 6.03       \\
1800               & Bias  (\%)  & BZ-$\mathcal{DP}$ & 26.98       & 46.85      & 76.99      &  & 9.16       & 5.54       & 1.15       \\
                  &                  & BMZ & 6.14        & 21.86      & 34.30      &  & 3.56       & 7.52       & 4.66       \\
                  &                  & Frailty & -117.34     & -154.61    & -237.30    &  & -77.21     & -101.53    & -110.87    \\
                  \cmidrule(lr){4-6}\cmidrule(lr){7-10}
                  &                  & BMZ-$\mathcal{DP}$ & 79.38       & 85.58      & 81.44      &  & 94.84      & 93.81      & 96.91      \\
                  &                  & BM-$\mathcal{DP}$ & 1.69        & 1.12       & 1.12       &  & 85.13      & 79.73      & 28.38      \\
                  & Coverage  (\%)        & BZ-$\mathcal{DP}$ & 49.48       & 30.93      & 21.64      &  & 82.47      & 86.59      & 94.84      \\
                  &                  & BMZ & 76.28       & 68.04      & 60.82      &  & 79.38      & 86.59      & 84.54      \\
                  &                  & Frailty & 1.69        & 1.69       & 1.69       &  & 84.75      & 72.88      & 40.68  \\
                  \bottomrule
\end{tabular}}
\end{table}

Based on the results in Table \ref{table:result}, the proposed BMZ-$\mathcal{DP}$ model achieves the overall best performance under all different sample sizes, with the smallest or among the smallest percentage bias and closest to 95\% coverage for all parameters. Specifically, under the proposed BMZ-$\mathcal{DP}$ model, the coverage probabilities for the survival process parameters are generally at the nominal level, whereas the coverage probabilities for the recurrent event parameters are slightly lower than nominal (though still the closest to nominal among all competing methods). With this level of sample size, the under-coverage for recurrent event model parameters is anticipated, due to the complicated event structure in the recurrent process including zero-inflation as well as nested random effects. In addition, BMZ-$\mathcal{DP}$ outperforms BM-$\mathcal{DP}$ particularly in uncovering the recurrent process by precisely tracking zero-inflation among the population. Meanwhile, we observe substantial estimation bias from both BZ-$\mathcal{DP}$ and  BMZ, as they either ignore the between-participant clustering or assume a fully parametric specification of the practice-level random effect. The performance of the frequentist joint frailty model is unsatisfactory in estimating both recurrent and survival related parameters, which is expected due to its omission for heterogeneity and misspecification of the distributions of the practice-level random effect and residual error of the survival process. \textcolor{black}{Finally, the results for the proposed BMZ-$\mathcal{DP}$ and its variations, BM-$\mathcal{DP}$,BZ-$\mathcal{DP}$ and BMZ, in Web Tables 2 and 3 are qualitatively similar to those in Table \ref{table:result}, suggesting that the proposed model is relatively robust when the data generation is simpler with single-component normal practice-level random effect, and when there exists individual-level heterogeneity regarding the terminal event process.}

\textcolor{black}{Overall, the difference in the performance of our method and the competing approaches across different scenarios we have investigated helps reinforce the necessity of the key components of our proposed model.} BM-$\mathcal{DP}$ and the frequentist joint frailty model carry the largest estimation bias and lowest coverage for the recurrent process parameters, indicating that accurate inference for recurrent event processes depends critically on adjusting for population heterogeneity. Similarly, BZ-$\mathcal{DP}$ also suffers with large bias with low coverage, suggesting the necessity for explicitly accounting for the multi-level data structure with our complex survival outcomes. Finally, the performance of BMZ demonstrates that it is critical to consider a flexible nonparametric prior for the practice-level random effect, as the estimation of model parameters can otherwise be subject to bias. Of note, we have also conducted a similar set of simulations when the baseline hazard for the recurrent process is generated from a Weibull distribution and the conclusions are essentially no different. The details of the simulation setting, model implementation and results are provided in Web Appendix S2.


\section{Application to the STRIDE Cluster Randomized Trial}\label{sec:data}

\subsection{Strategies to Reduce Injuries and Develop Confidence in Elders (STRIDE) Trial}


\textcolor{black}{As stated in Section 1, STRIDE was a pragmatic, parallel CRT aimed at reducing serious falls among community-dwelling older adults. A total of 5,419 participants aged 70 years and older from 86 primary care practices are included in our final analysis. Primary care practices range in size from 10 to 158 participants, with a mean cluster size 63 and coefficient of variation 0.52. The participants were followed for a maximum of 44 months, at which point the survival outcomes were right censored due to study termination. During the follow-up, each occurrence of fall injury and its severity level, along with adverse events including hospitalization and death, were recorded periodically. The descriptive statistics summarizing the number of recurrent adjudicated serious fall injuries, observed death events, and key baseline covariates are presented in Web Table 4 by arm. In the intervention and control primary care practices, the event rate (first serious fall-related injury) was 5.2 and 4.9 per 100 person-years of follow-up, respectively, while in both treatment arms the death rate was lower, at 3.4 per 100 person-years of follow-up. Under cluster randomization, the baseline characteristics were generally balanced between arms, although slightly more white elderly patients appeared in the intervention practices, and patients from the intervention practices tended to have slightly more chronic disease conditions at baseline.} 

\subsection{Model Specification and Implementation}\label{sec:implement}

\textcolor{black}{We implement our Bayesian semi-parametric joint model and the competing approaches described in Section \ref{sec:simu} to analyze the STRIDE cluster randomized trial, investigating how the intervention and covariates are associated with serious fall injuries and death among the elderly participants who are 70 years of age or older. We are interested in the recurrent adjudicated serious fall injuries (falls resulting in a fracture, joint dislocation, cut requiring closure, or overnight hospitalization, reported by participants and confirmed by medical records or claims data) and deaths. Besides the intervention, we adjust for several risk factors, including age, sex, race, and number of chronic coexisting conditions (NCD), to study their effect on fall prevention and survival. We implement the proposed BMZ-$\mathcal{DP}$ model and the competing approaches, i.e. BZ-$\mathcal{DP}$, BM-$\mathcal{DP}$, BMZ as well as the joint latent class model developed by \citet{xu2021joint}, assuming all the risk factors can potentially impact the recurrent events and survival. In particular, the joint latent class model does not account for zero inflation nor clustering by practice; however, it allows for latent-class-specific effect on recurrent event and survival. To implement the first four approaches, we set $\sigma^2_{\zeta}=\sigma^2_{\xi_1}=\sigma^2_{\xi_2}=10$ as prior variances for the regression parameters. For the $\mathcal{DP}$ prior of the practice-specific random effects, we set $\sigma^2=1$ for the normal base measure. We further set $a_{\kappa}=b_{\kappa}=1$ for the Gamma base measure for $\mathcal{DP}$ prior of the participant-specific shape parameter $\kappa_{ij}$. We assign $\mathcal{IG}(1/2, 1/2)$ as conjugate hyper-priors for the prior variances $\sigma^2_{\beta}$, $\sigma^2_{\alpha}$ and $\tau_j$. For the recurrent event baseline hazard function, we consider $G=5$ quantile grids $\bfs$ and an improper prior for each element of $\bflambda$ as indicated in Section \ref{sec:prior}. Other hyper-parameter specifications closely follow those in Section \ref{sec:simu}. For $p_{ij}$, besides assigning $p_{ij}=0.5$, we also consider a logistic model to represent $p_{ij}$ to allow for dependence on treatment, sex and race, which accommodates the potential effect of intervention and baseline risk factors on the individual susceptibility status. We compare the results under both prior settings as discussed in Section \ref{sec:SA}. For the joint latent class model, we consider the default implementation and mixture of finite mixtures hierarchical prior for latent class probability explained in \citet{xu2021joint}. For each model, under random initials, we run MCMC for 20,000 iterations with the first 10,000 as burn-in. The trace plots for several key parameters are provided in the Web Figure 2. Finally, the posterior results for each method are summarized in Table \ref{table:realdata}.  Of note,  the joint frailty model under a frequentist paradigm (based on the \texttt{frailtypack}) is not included given the model did not converge after running for $48$ hours.}

\begin{table}[htbp]
 \begin{center}
 \caption{Posterior inference results for parameters in the recurrent and survival processes under different methods for the analysis of the STRIDE study.}  \label{table:realdata}
\resizebox{\textwidth}{!}{
  \begin{tabular}{cccccc}
  \toprule  
  & BMZ-$\mathcal{DP}$ & BM-$\mathcal{DP}$ & BZ-$\mathcal{DP}$& BMZ& \citet{xu2021joint} \\
  \midrule 
  \multicolumn{6}{c}{Recurrent Process}\\
  \midrule  
  Intervention  & -0.08 (-0.34, 0.16) & -0.25 (-0.48, 0.01) & -0.02 (-0.26, 0.22) & -0.12 (-0.28, 0.04)&-0.34 (-0.57, -0.11) \\
  NCD & 0.14 (0.06, 0.22) & 0.14 (0.08, 0.18)&  0.14 (0.07, 0.22)& 0.10 (-0.04, 0.24)&-0.22 (-0.29, -0.15)\\
  Age & -0.05 (-0.06, -0.04) & -0.05 (-0.06, -0.04) & 0.01 (-0.01, 0.01) & -0.04 (-0.09, -0.04)&-0.02 (-0.24, 0.20)\\
  Sex (Female) & 0.08 (-0.12, 0.30) & 0.03 (-0.12, 0.20) &  0.18 (-0.01, 0.36) & -0.02 (-0.18, 0.14)&0.08 (-0.02, 0.18)\\
  Race (White) &  0.18 (-0.18, 0.47) & 0.17 (-0.13, 0.44) & 0.81 (0.38, 1.29) & 0.89 (0.48, 1.30)&0.11 (-0.43, 0.65)\\
  \midrule 
  \multicolumn{6}{c}{Survival Process}\\
   \midrule  
  Intervention &  0.02 (-0.23, 0.24) & -0.05 (-0.20, 0.11) &  -0.04 (-0.38, 0.28) & 0.11 (-0.16, 0.38)&-0.10 (-0.34, 0.14)\\
  NCD & -0.16 (-0.26, -0.10) & -0.15 (-0.21, -0.12) & -0.19 (-0.31, -0.09) & -0.18 (-0.30, -0.05)&-0.26 (-0.33, -0.19)\\
  Age &-0.03 (-0.06, -0.01) & -0.04 (-0.05, -0.03) & -0.04 (-0.07, -0.01) & 0.02 (0.00, 0.04)&0.10 (-0.03, 0.23)\\
  Sex (Female)  & 0.50 (0.22, 0.81) & 0.45 (0.32, 0.61) & 0.65 (0.29, 1.16) & 0.97 (0.50, 1.50)&1.17 (0.85, 1.49)\\
  Race (White) &0.06 (-0.45, 0.52) & 0.09 (-0.08, 0.30) &0.04 (-0.69, 0.62) & 0.12 (-0.33, 0.67) &0.19 (0.07, 0.31)\\
 \midrule 
  {LPML} &  -1534.12 & -3585.48 & -1624.85 & -2092.51&-3466.78\\
  \bottomrule
  \end{tabular}}
  \end{center}

\end{table} 

\subsection{Results from the Proposed Model}\label{sec:results_main}

\textcolor{black}{We first investigate the impact of the intervention and risk factors on recurrent fall injuries. Under the proposed BMZ-$\mathcal{DP}$ model, NCD and age appear to be associated with fall injury intensity, and their corresponding 95\% credible intervals exclude zero. Exponentiating 
the posterior means of the model parameters, we find that one additional chronic condition multiplies the serious fall rate by 1.15 
and a one year increase in age reduces injury intensity by around 5\%.
The latter result provides seemingly counter-intuitive evidence since it was originally believed that older age increases the risk for 
fall injury. 
However, because the population recruited in our study are 70 years of age or older, it is also likely that a further increase in age could start to prevent them from potential triggers for serious fall such as exercise or intensive movement. In addition, female and white patients are more likely to experience falls than other subgroups. On the hazard scale, the intervention appears to have a small effect on reducing the risks for recurrent fall injuries. For the terminal event---death, NCD, age and sex remain significant predictors, with their 95\% credible intervals excluding zero. An increase in NCD or age leads to shorter survival times, as expected, but female and white patients appear to have longer survival. 
Consistent with the previous analysis focusing on the first occurrence of fall injury \citep{stride_introduction,li2022comparison,chen2022finite}, the intervention reduces the rate of recurrent falls as well as benefits survival experiences, but 95\% credible intervals of the intervention effect parameter in both processes include zero.}

\begin{figure}[ht]
\centering
     \includegraphics[width=1.0\textwidth]{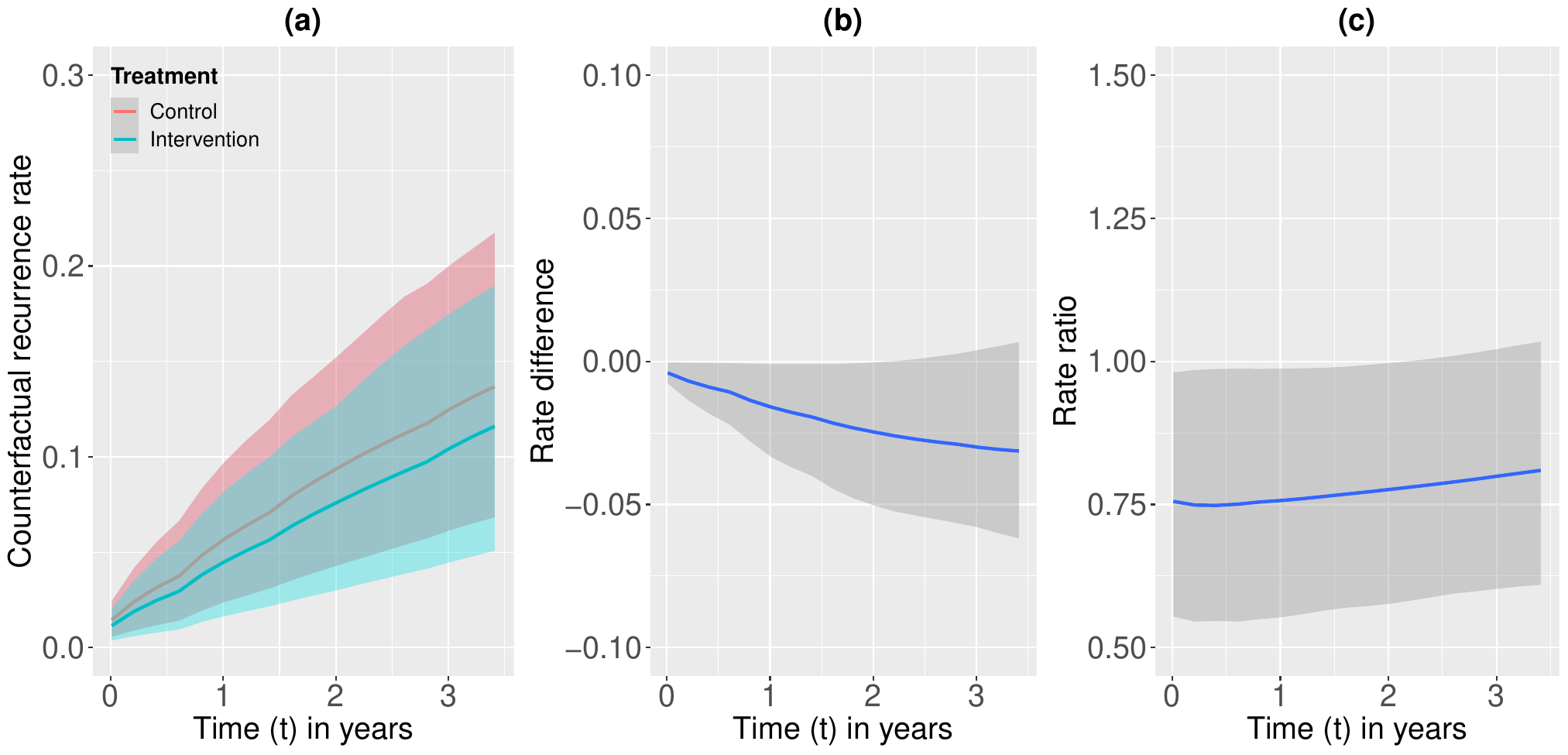}
    \caption{
    (a) Posterior mean and 95\% pointwise credible bands for counterfactual recurrence rates over time; (b) posterior mean and 95\% pointwise credible bands for the counterfactual recurrence rate difference over time; (c) posterior mean and 95\% pointwise credible bands for the counterfactual recurrence rate ratio over time.}\label{fig:curve}
\end{figure}

\textcolor{black}{To further interpret the treatment effect on the risk of fall-injury, we borrow the counterfactual outcome framework to investigate the participant-average treatment effect \citep{kahan2022estimands}. Specifically, suppose an individual remains alive at time $t$, the recurrence rate for that individual at time $t$ is given by $\mu_{ij}(t)=E[N_{ij}(t)]=p_{ij}\int_0^t\lambda_{ij}(u)H_{ij}(u)du$, where we recall $p_{ij}$ is the participant-specific probability to be classified into the unsusceptible subgroup (therefore not a structural zero), $\lambda_{ij}(t)$ is the participant-specific hazard function for the recurrent event, and $H_{ij}(t)$ is the survival function for the terminal event; similar definition has also bee discussed in \citet{xu2021joint} in the absence of zero-inflation. Then the counterfactual recurrence rate had an individual received the intervention (possibly contrary to fact) can be expressed by $\mu_{ij}(1,t)=p_{ij}(do(\text{Treat}=1))\int_0^t\lambda_{ij}(u|do(\text{Treat}=1))H_{ij}(u|do(\text{Treat}=1))du$, where we use the $do$-calculus notation to indicate the critical step of setting the treatment variable to be $1$ when computing the probability to be in the unsusceptible subgroup, recurrent event hazard and terminal event survival functions \citep{10.5555/331969}. Analogously we define $\mu_{ij}(0,t)=p_{ij}(do(\text{Treat}=0))\int_0^t\lambda_{ij}(u|do(\text{Treat}=0))H_{ij}(u|do(\text{Treat}=0))du$. Therefore the participant-average treatment effect for fall injury at time $t$, on the rate difference scale and rate ratio scale, can be expressed as
\begin{align*}
\text{rate difference}(t)=\frac{\sum^J_{j=1}\sum^{N_j}_{i=1}\{\mu_{ij}(1,t)-\mu_{ij}(0,t)\}}{\sum^{J}_{j=1}N_j},~~~
\text{rate ratio}(t)=\frac{\sum^J_{j=1}\sum^{N_j}_{i=1}\mu_{ij}(1,t)}{\sum^J_{j=1}\sum^{N_j}_{i=1}\mu_{ij}(0,t)}.
\end{align*}
Figure \ref{fig:curve} plots the posterior mean and 95\% credible intervals on the counterfactual recurrent rates, rate difference and rate ratio as a function of follow-up time in years. Panel (a) suggests a very mild treatment effect since intervention leads to a slightly lower counterfactual recurrence rate. Indeed, from panels (b) and (c) the recurrence rate difference at year 3 is around $-0.029$, suggesting around $29$ falls prevented per $1000$ patients; the recurrence rate ratio at year 3 is around $0.80$. Interestingly, the 95\% pointwise credible bands just exclude null for both the rate difference and rate ratio effect measures since the start of follow-up until approximately year $2$, but include the null from that time onward. It is important to note that these counterfactual treatment effect quantities are not identical to the treatment effect parameter due to non-collapsibility, and may be more interpretable when the interest lies in measuring the population impact of intervention due to switching from usual care to the fall injury prevention program.}

\begin{figure}[htbp]
\centering
\includegraphics[width=0.8\textwidth]{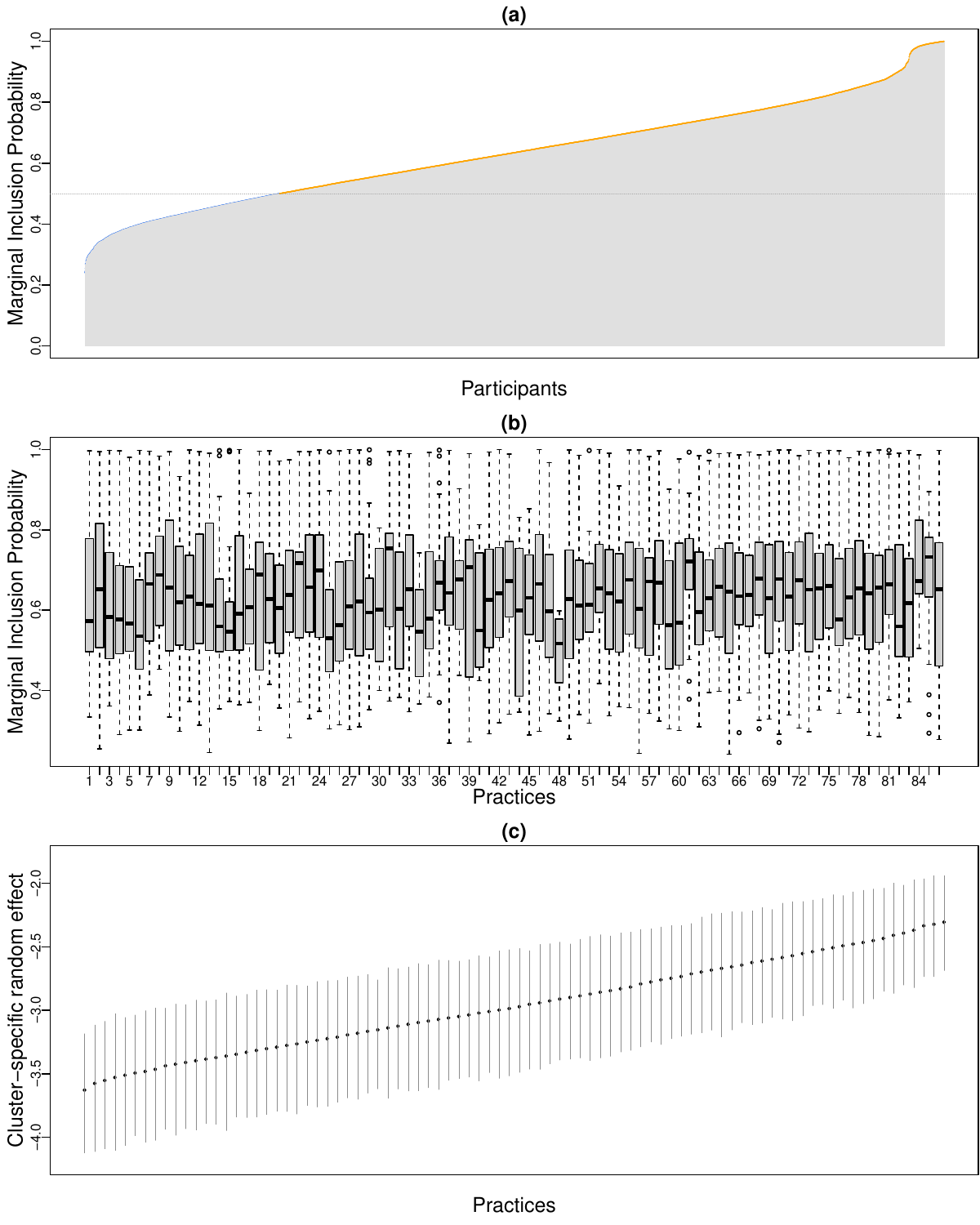}
\caption{Analysis results on structural zero probabilities and cluster-specific random effects for STRIDE: panel (a) displays the marginal posterior inclusion probability of $D_{ij}=1$ ordered from smallest to largest; panel (b) summarizes the distribution of marginal posterior inclusion probability of $D_{ij}=1$ within different practices; panel (c) provides the posterior inference of $\mu_j$ ordered by posterior mean.  }\label{realdataresult}
\end{figure}

Finally, we examine the structural zero probabilities and cluster-specific random effects, as these are two key aspects of STRIDE our method is well-suited to identify. We summarize the marginal posterior inclusion probability for $D_{ij}=1$ over subjects who may be considered naturally unsusceptible for the recurrent fall events in panel (a) of Figure \ref{realdataresult}. With a 0.5 cutoff 
\citep{barbieri2004optimal}, we conclude there could be a substantial fraction of participants who may not be susceptible to serious fall injuries during the study period. We present the distribution of those inclusion probabilities within different practices in panel (b) of Figure \ref{realdataresult}. Almost all practices include a large amount of patients from the unsusceptible subgroup, and the inclusion probability varies both within and between practices. This visualization can help identify practices with substantially more unsusceptible patients for falls, although the exact scientific mechanism for unsusceptibility remains to be further studied. We also provide the posterior mean along with the 95\% credible intervals obtained from the inference of each $\mu_j$ in panel (c) of Figure \ref{realdataresult}, where the practices are ordered by their point estimates. Clearly, the cluster-level frailties show substantial heterogeneity across practices, are all negative, and are all significantly different from zero, so failure to account for this between-practice heterogeneity will likely result in bias in the association estimates.

\subsection{Model Comparison and Sensitivity to Priors}\label{sec:SA}

\textcolor{black}{Table \ref{table:realdata} additionally summarizes results from the competing models. Although results from the models without zero-inflation (BM-$\mathcal{DP}$), multi-level data structure (BZ-$\mathcal{DP}$) or both DP priors (BMZ) are generally consistent with those from BMZ-$\mathcal{DP}$, we notice differences in the effect estimates and credible intervals. For example, the main conclusions based on BM-$\mathcal{DP}$ align with those from BMZ-$\mathcal{DP}$, but failure to account for the structural zeros tends to exaggerate the intervention effects for the serious fall intensity and the survival. The results from BZ-$\mathcal{DP}$ did not identify age as an important predictor for recurrent fall injury, and the BMZ model results also reveal larger intervention effect estimates than the proposed model. Finally, the joint latent-class model ignores both the zero-inflation and clustering, and tends to overestimate the covariate and intervention effects for both recurrence and survival processes.}

\textcolor{black}{To evaluate model fit, we consider model validation diagnostics for recurrent events data using the conditional predictive ordinate (CPO); the $\text{CPO}$ refers to the conditional predictive ordinate used for detecting surprising observations over subjects \citep{cpo_introduction_2}, and has similarly been operationalized in \citet{sinha2008current} for non-clustered recurrent event data. By plotting $\text{CPO}_{ij}$ against the observed follow-up time $\widetilde{R}_{ij}$, we can visually assess whether the posterior analysis has any unusually low prediction capabilities for certain values of the observed survival time. Web Figure 3 indicates lack of association between $\text{CPO}_{ij}$ and $\widetilde{R}_{ij}$ under the proposed model, therefore suggesting no evidence against the model adequacy. In addition, we compare the proposed model with the competing models by calculating the log pseudo marginal likelihood (LPML) based on leave-one-out-cross-validation as $\text{LPML}=\sum_{j=1}^J\sum_{i=1}^{N_j} \log(\text{CPO}_{ij})$. This is a commonly used metric in Bayesian survival analyses to compare model performance, with a larger value suggesting a better fit to the data. As shown in Table \ref{table:realdata}, our proposed model has the largest LPML and therefore demonstrates the best fit to the analysis of STRIDE data among the competing models.}


\begin{table}[htbp]
 \begin{center}
\caption{Posterior inference for parameters under the proposed model but with different prior specifications. The second column represents the primary implementation in Section \ref{sec:implement}, and the remaining columns correspond to sensitivity analyses under the following scenarios: (a) prior for marginal inclusion independent of covariates; (b): alternative hyper-priors for the prior variances of the regression coefficients; (c): expanding quantile grids for the recurrent event baseline hazard; (d) slightly more informative priors for frailty coefficients and logistic structural-zero regression coefficients; (e) less informative hyper-prior for practice-specific variance parameters.}
 \label{table:sensitive}
\resizebox{\textwidth}{!}{
  \begin{tabular}{ccccccc}
  \toprule 
  & Section \ref{sec:implement} &(a)& (b)&(c)&(d)&(e)\\
  \midrule 
  \multicolumn{7}{c}{Recurrent Process}\\
  \midrule  
  Intervention  & -0.08 (-0.34, 0.16) & -0.12 (-0.28, 0.04)& -0.03 (-0.22, 0.17) &-0.01 (-0.25, 0.23)&-0.05 (-0.25, 0.17)&-0.06 (-0.38, 0.20)\\
  NCD & 0.14 (0.06, 0.22) & 0.10 (-0.04, 0.24)& 0.11 (0.05, 0.19)&0.12 (0.04, 0.19)&0.11 (0.04, 0.19)&0.13 (0.05, 0.23)\\
  Age & -0.05 (-0.06, -0.04) & -0.04 (-0.09, -0.04)& -0.05 (-0.06, -0.04)&-0.05 (-0.06, -0.04)&-0.05 (-0.07, -0.04)&-0.04 (-0.05, -0.04)\\
  Sex (Female) & 0.08 (-0.12, 0.30) &  -0.02 (-0.18, 0.14)& 0.06 (-0.11, 0.25)&0.02 (-0.17, 0.24)&0.05 (-0.18, 0.26)&0.07 (-0.13, 0.26)\\
  Race (White) &  0.18 (-0.18, 0.47) &  0.89 (0.48, 1.30)& 0.21 (-0.21, 0.56)&0.31 (-0.11, 0.69)&0.16 (-0.27, 0.50)&0.26 (-0.13, 0.82)\\
  \midrule
  \multicolumn{7}{c}{Survival Process}\\
   \midrule  
  Intervention & 0.02 (-0.23, 0.24) & 0.11 (-0.16, 0.38)& -0.03 (-0.34, 0.29)&0.04 (-0.29, 0.33)&-0.01 (-0.29, 0.23)&0.05 (-0.22, 0.34)\\
  NCD & -0.16 (-0.26, -0.10) & -0.18 (-0.30, -0.05)& -0.18 (-0.28, -0.07)&-0.17 (-0.28, -0.08)&-0.15 (-0.24, -0.08)&-0.14 (-0.23, -0.07)\\
  Age & -0.03 (-0.06, -0.01) &  0.02 (0.00, 0.04) &-0.03 (-0.05, -0.01)&-0.03 (-0.05, -0.02)&-0.01 (-0.02, 0.01)&-0.01 (-0.02, -0.00)\\
  Sex (Female)  & 0.50 (0.22, 0.81) &  0.97 (0.50, 1.50)& 0.62 (0.30, 0.98)&0.58 (0.27, 0.95)&0.52 (0.25, 0.81)&0.44 (0.20, 0.72)\\
  Race (White) &  0.06 (-0.45, 0.52) &  0.12 (-0.33, 0.67) & 0.00 (-0.54, 0.49)&0.31 (-0.11, 0.69)&0.09 (-0.36, 0.53)&-0.07 (-0.61, 0.38)\\
 \midrule 
  {LPML} & -1534.12 & -2092.51& -1580.62&-1554.66&-1689.87&-1527.97\\
  \bottomrule
  \end{tabular}}
  \end{center}
\end{table} 

\textcolor{black}{Finally, we perform a series of sensitivity analyses to evaluate how much the posterior inference results would change according to alternative prior specifications. We independently check the estimation results under the following scenarios. (a) The hyper-prior for the marginal inclusion indicator is uniformly set to $p_{ij}=0.5$ without dependence on covariates. (b) The hyper-priors for the prior variances associated with the regression coefficients are set to be $\sigma_{\beta}^2\sim \mathcal{IG}(0.01,0.01)$ and $\sigma_{\alpha}^2\sim\mathcal{IG}(0.01,0.01)$ instead of $\mathcal{IG}(1/2,1/2)$, so they are less informative. (c) The number of quantile grids $\bfs$ associated with the baseline hazard in the recurrent event process is expanded to $G=8$ (rather than $G=5$ in our original implementation), and we specify a proper uniform prior $(0, 100)$ for each element within $\bflambda$ (rather than the improper prior in our original implementation). (d) Setting $\sigma^2_{\zeta}=\sigma^2_{\xi_1}=\sigma^2_{\xi_2}=3$ to give slightly more informative priors for the frailty association parameters and the logistic structural-zero regression parameters. (e) Setting the hyper-prior for the practice-specific variance parameter to be $\tau_j\sim \mathcal{IG}(0.01,0.01)$ to make it less informative. The posterior inference results for each specification are summarized in Table \ref{table:sensitive}, and we observe that the results are generally stable across different specifications. We further obtain the LPML for each analysis and found that only specification (e) produces slightly larger LPML than (but very close to) our primary implementation in Section \ref{sec:implement}. The posterior inference results for the regression coefficients are not substantially different between these two specifications, and therefore still support our interpretation in Section \ref{sec:results_main}.}

\section{Discussion}\label{s:discuss}
In this paper, we propose a new Bayesian semi-parametric joint model framework to simultaneously characterize the recurrent event process and survival in the presence of clustering and potential zero-inflations. To accommodate the between-participant clustering commonly seen in pragmatic clinical trials, we introduce hierarchical random effects at the participant and practice levels, both of which bridge the recurrent and survival processes. To further relax the parametric assumptions, we specify separate nonparametric realizations for the baseline hazard, practice-level random effect and terminal event survival function, which are naturally incorporated into our unified Bayesian paradigm and enhance the model robustness compared with the existing alternative modeling strategies. We also demonstrate the necessity of each analytical component within our joint modeling framework and develop MCMC algorithms to enable posterior inference for all model parameters. Through extensive simulations and our application to a recent pragmatic cluster randomized trial STRIDE, we demonstrate the advantage of our method by providing more reliable estimation and inference while maintaining robustness to violations of certain parametric assumptions. 

To account for zero-inflation, we define the structural zeros as those contributed by participants who are unsusceptible to the recurrent event during the study period, and we impose a two-component mixture to represent the recurrent event hazard. This definition is similar to the existing cure model literature \citep{yu2004joint,rondeau2013cure} with straightforward interpretation of unsusceptibility. When recurrent events are subject to a terminal event, alternative definitions of unsusceptibility exist. For example, one can either define the unsusceptibility status based on the unsusceptibility status of recurrent events or further differentiate the structural zero from the random zero based on the terminal event. The former definition has been adopted in \cite{xu2018joint} and \cite{han2020variable} as well as our current model specification, given study participants should always be susceptible to typical terminal events such as death, particularly in studies with an elderly population and a longer follow-up. \cite{liu2016joint} also provided a brief explanation on the difference of these two definitions of unsusceptibility driven by a specific application. In any case, our modeling framework can be easily adapted under an alternative definition of unsusceptibility incorporating the terminal event.

One potential limitation of our modeling strategy is that we have not distinguished the survival hazard between subjects who experience recurrent event(s) and those who directly move to the terminal event. The impact upon the survival function from the previous recurrent event occurrence is primarily controlled by the shared hierarchical frailty terms, $\gamma_{ij}$ and $\mu_j$, from the two hazards. An alternative modeling strategy can be based on the multi-state model \citep{lee_2016,li2022comparison}, where one can more explicitly split the event occurrence paths into 1) recurrent events only, 2) recurrent events followed by the terminal event, and 3) terminal event only, and characterize state-specific hazard functions for each of them in conjunction with different random effects. Under multi-state modeling, we may be able to capture the influence from both fixed and random effects on different hazard functions, potentially allowing for richer information extraction. However, a practical issue of multi-state modeling in our case comes from the growing number of unknown parameters, since we would need an additional set of coefficients as well as more random effects. This could induce computational and inferential challenges, especially with a small sample size, and merits additional investigation. \textcolor{black}{Another potential limitation is that we have not distinguished the terminal event survival processes for the susceptible and unsusceptible patient subpopulations. In STRIDE, It is possible that patients who are unsusceptible to recurrent fall injuries are more likely to survive until the end of the study that those who are susceptible. Generally, this assumption is challenging to test empirically from data because in our model formulation these two subpopulations are latent (therefore not fully observed from the study sample alone). A possible improvement of the proposed model is to modify the likelihood in Section \ref{sec:posterior} to allow the terminal event survival function and density function to further depend on $D_{ij}$. However, the associated posterior inference can become substantially more challenging with our nonparametric prior specifications at multiple levels. Similarly, a primary care practice including more than usual unsusceptible patients may be expected to have lower risk of death, and it would be interesting to further control for $N_j^{-1}\sum_{i=1}^{N_j}D_{ij}$ in the terminal event model. Although these relevant extensions are promising for understanding the complex event processes in STRIDE, the identifiability and efficient posterior inference based on these extensions are beyond the scope of this article and will be pursued in future research.}


\section*{Supplementary Materials}
Detailed MCMC algorithms and additional results are available as supplementary materials. Sample code to implement the proposed Bayesian model is available at \url{https://github.com/xt83/Bayesian_semi_parametric_inference_for_clustered_recurrent_event}.

	\bibliographystyle{asa}
\baselineskip=10pt
\bibliography{biomsample}



\end{document}